\documentclass[prd,superscriptaddress,showpacs,preprintnumbers,amsmath,amssymb]{revtex4}
\usepackage{amsmath}
\usepackage{amstext}
\usepackage{amsopn}
\usepackage{amsfonts}
\usepackage{amssymb}
\usepackage{bbm}
\usepackage{accents}
\usepackage{empheq}
\usepackage{graphicx}
\usepackage{epsf}
\usepackage{graphics}
\usepackage[latin1]{inputenc}
\def\sc{\scriptscriptstyle}
\begin{document}

\title{Motion of a test particle  in the transverse space of D$p$-branes}
\author{Anindita Bhattacharjee}
\affiliation{Department of Physics, Assam University, Silchar, Assam 788011, India}
\author{Ashok Das}
\affiliation{Department of Physics, University of Rochester, Rochester, NY 14627, USA}
\affiliation{Saha Institute of Nuclear Physics, 1/AF Bidhannagar, Calcutta 700064, India}
\author{Levi Greenwood}
\affiliation{Department of Physics, University of Rochester, Rochester, NY 14627, USA}
\author{Sudhakar Panda}
\affiliation{Harish-Chandra Research Institute, Chhatnag Road,
Jhusi, Allahabad-211019, India}

\begin{abstract}
We investigate the motion of a test particle in higher dimensions due to the presence of extended sources like D$p$-branes by studying the motion in the transverse space of the brane. This is contrasted with the motion of a point particle in the Schwarzschild background in higher dimensions. Since D$p$-branes are specific to $10$-dimensional space time and exact solutions of geodesic equations for this particular space time has not been possible so far for the Schwarzschild background, we focus here to find the leading order solution of the geodesic equation (for motion of light rays).  This enables us to compute the bending of light in both the backgrounds. We   show that contrary to the well known result of no noncircular bound orbits for a massive particle, in Schwarzschild background, for $d\geq 5$, the $Dp$-brane background does allow bound elliptic motion only for $p = 6$ and the perihelion of the ellipse regresses instead of advancement. We also find that circular orbits for photon are allowed only for $p\le 3$.

\end{abstract}
\maketitle

\section{Introduction}
In classical theory of general relativity we study the gravitational effects of massive point particles (since far away from the source the gravitating object can be thought of as a point particle) and there are various tests of general relativity based on the analysis of the motion of point test particles (both massive and massless) in such a background. For a variety of reasons there has been substantial progress, over the years, in understanding general relativity in higher dimensional space-times. In fact, string theory is one of the leading candidates for the ultraviolet complete theory of gravity and the consistency of  the super string theory requires that our universe should have nine spatial dimensions in contrast  to the three spatial dimensions of the presently observed universe. This discrepancy  is best resolved by the proposal of compactification which says that the extra six spatial directions are compact and the size of this compact space is so small that present experiments cannot possibly probe it directly. On the other hand, it is quite likely that the low energy theory would capture some of the effects of compactification from higher dimensions which can possibly be tested at LHC. From the analysis of a toy model of compactification from five to four dimensions, it is becoming overwhelmingly clear that in such a scenario, not only will the standard model particles be duplicated  but the Kaluza-Klein modes of some of these particles will also have clear signals in experiments at LHC \cite{bdmr}. While we await the confirmation (or absence) of such signals from the accelerator experiments, it is worth analyzing large scale tests of higher dimensional gravity. In fact, there already exist substantial literature on (higher dimensional) Kaluza-Klein gravity \cite{OW}. This theory can be tested by analyzing the motion of test particles in the background of a static, spherically symmetric mass like the sun. The generalization of Birkhoff's theorem  to higher dimensional compactified theories \cite{BM},\cite{HJ} allows us to construct static solutions for the metric and in the five dimensional case most of the attention has been focussed thus far on the solitonic metric \cite{GP},\cite{RD},\cite{DO} which satisfies the five dimensional vacuum field equations and which reduces to the standard Schwarzschild solution on a four dimensional hyper-surface without explicit dependence on the fifth dimension. 

The motion of test particles in the gravitational background of the soliton has been studied in connection with the well known classical tests of general relativity, e.g. gravitational redshift, light deflection, perihelion advance, time delay etc \cite{OLD},\cite{KWE},\cite{LO}.  For the five dimensional case, the soliton metric is of the form
\begin{equation}
ds^2 = A^a dt^2 - A^{-a -b} dr^2 - A^{1-a-b} r^2 (d\theta^2 + \sin^2 \theta d\phi^2) -A^b dy^2,\label{1}
\end{equation}
where $y$ denotes the coordinate of the fifth dimension and $ A(r) = 1 - 2M/r$. Here,  $M$ is the parameter (including the gravitational constant) related to the mass of the soliton taken to be at the origin of the coordinate system and the exponents $a, b$ are constant parameters satisfying $ a^2 + a b + b^2 = 1$. Thus the metric has two independent parameters. Note that we can recover the standard four dimensional Schwarzschild metric on the hyper-surface ($y =$ constant) in the limit $a=1, b = 0$. Hence $b$ can be considered as the meaningful free parameter of the theory in the sense that a non-zero value of $b$ signals a departure from the Schwarzschild  geometry of a massive point gravitating source.

While the above analysis is motivated from the point of view of  Kaluza-Klein gravity in five dimensions, we will be focussing in the following on the case of (super) string theory which is consistent in ten dimensions. This theory contains  point like excitations in its perturbative spectrum as well as extended but stable objects called D$p$-branes in its non-perturbative spectrum\cite{pol}. The D$p$-branes define hypersurfaces in the ten dimensional space-time with $p$-spatial and one time like coordinates. Therefore, they are $p+1$ dimensional objects with $9-p$ transverse spatial directions and in this theory it will be of interest to study the gravitational effects, not only of massive point like sources but also of massive branes i.e. extended objects, on probe particles moving in the space transverse to the brane. This would allow us to compare and distinguish the effects of point like sources from those of sources which are extended in nature e.g. strings, membranes or other branes. 

Therefore, we are going to study the $10$-dimensional curved space-time metric obtained in string theory describing such objects.  Let us consider $N$ coincident D$p$-branes described by the metric ($c=1$) \cite{duff} 
\begin{equation}
ds^2 = d\tau^{2} = H^{-\frac{1}{2}}\left(dt^2-\sum_{i=1}^p (dx^i)^2\right)-H^{\frac{1}{2}}\left(dr^2+r^2 d\Omega^2_{8-p}\right),\label{2}
\end{equation}
where $r$ denotes the radial coordinate and $d\Omega^{2}_{8-p}$ the angular element of the $(9-p)$-dimensional transverse space. The harmonic function $H (r)$ is given by
\begin{equation}
H(r)=1+\frac{Q_p}{r^{7-p}},\label{3}
\end{equation}
where $Q_p = d_p N g_s \ell_s^{7 - p} $ with $g_{s}, \ell_{s}$ denoting respectively the string coupling and string length scale  and $d_p = 2^{5-p} \pi^{(5 - p)/2} \Gamma ((7 - p)/2)$. For all practical purposes we can assume that $Q_{p} \sim \ell_{s}^{7-p}$.  We note here that
\begin{equation}
H (r)\xrightarrow{\text{large}\ r} 1,\quad H (r) \xrightarrow{\text{small}\ r} \frac{Q_{p}}{r^{7-p}},\label{3a}
\end{equation}
where large and small $r$ are defined with respect to the string length $\ell_{s}$. The metric \eqref{2} satisfies  the Einstein equation derived from the $10$-dimensional gravitational action in the presence of a Maxwell term with a field strength of rank $p+2$ (for notations and other details see \cite{duff}). We will study the behavior of a point test particle moving under the influence of this gravitational background in the space transverse  to the $p$-brane. Let us note that the form of the line element given in \eqref{2} is clearly different in character from the Schwarzschild  line element (to be discussed in next section) as well as the solitonic case in \eqref{1}, namely, the function $H^{\frac{1}{2}}$ multiplies both the radial and the angular line elements for the case of the branes which is to be contrasted with the other cases.  
Thus,  the analysis we will carry out in this paper will have a different character from the five dimensional case described above.

 We point out here that there has already been an extensive study of the motion of probe branes  in the transverse space of stacks of static branes of various kinds (see for example \cite{kut},\cite{kuta}, \cite{sen}, \cite{dpr1}, \cite{dpr2}, \cite{kp},\cite{ctt}) which have also been used in cosmological applications \cite{tw}, \cite{pst}, \cite{pstw}, \cite{ps},\cite{psh}. Also, in mirage cosmology \cite {ms},\cite{BS}, the motion of a D-brane is considered in a gravitational background. On the contrary,  it is the analysis of the motion of  probe particles which is the goal of the present investigation. In a sense, since higher dimensional gravitational theories do admit extended objects as gravitational sources, it is quite natural to investigate the motion of a probe point particle in the transverse space of branes (i.e. in the gravitational field produced by extended objects like branes) so as to contrast the results with those obtained from the motion of a point particle in the gravitational field produced by point sources.

Before we proceed further, it is useful to discuss briefly about the distance scales at which the gravitational effects of these branes become important. We recall that in a space-time of arbitrary dimension $d$, the gravitational potential created by a point particle of mass $M$ at a distance $r$ is given by 
\begin{equation}
V \simeq  - \frac{G^{(d)} M}{r^{d-3}},\label{4}
\end{equation}
where $G^{(d)}$ denotes Newton's gravitational constant in $d$-dimension. Working in natural units where $V$ is dimensionless, we define a characteristic gravitational length scale $L$ given by 
\begin{equation}
L^{d-3} \equiv G^{(d)} M,\label{4a}
\end{equation}
so that we can write $V = - (L/r)^{d-3}$. This shows that gravitational effects are weak for $r >> L$ but become important at scales of order $L$. Note that for a point gravitational source, $L$ coincides with the Schwarzschild radius of a black hole of mass $M$   in $d$-dimensional space-time (up to factors of order unity). 

When we have a stack of $N$ D$p$-branes wrapped around a (spatial) $p$-dimensional compact volume $V_p$, the effective mass of the branes is given by
\begin{equation} 
M = N T_p V_p \simeq \frac{N V_p}{g_{s} (\sqrt{\alpha^{\prime}})^{p + 1}},\label{5}
\end{equation}
where $T_p$ is the tension of the D$p$-brane and $\alpha^{\prime}$ corresponds to the slope of the Regge trajectory (which is related to the string tension or the square of the string length scale $\ell_{s}$). The dimensionally reduced space-time in this case is $(d - p)$-dimensional. In this space-time, the branes can be thought of as a point source of mass $M$. It follows from \eqref{4a} that the characteristic  size $L$ of such a system (where $p$ dimensions have been compactified) is given by \cite{barton}
\begin{equation}
L^{d-p - 3} = G^{(d- p)} M = \frac{G^{(d)} M}{V_p} = \frac{G^{(d)} N}{g_{s} (\sqrt{\alpha^{\prime}})^{p+1}}.\label{6}
\end{equation}
Using the relation between the Newton's constant in $d$-dimensions and the string coupling  
\begin{equation}
G^{(d)} \sim g_{s}^{2} (\sqrt{\alpha^{\prime}})^{d-2},\label{7}
\end{equation}
in \eqref{6} we obtain
\begin{equation}
L^{d - p - 3} \simeq g_{s} N (\sqrt{\alpha^{\prime}})^{d - p - 3}.\label{8}
\end{equation} 
Note from \eqref{8} that for $d = 10$ and for $p \leq 6$, we have $L \rightarrow 0$ as $g_{s}N \rightarrow 0$, i.e. the gravitational effects of the branes vanish. Thus, we need $g_s \rightarrow 0$ and a sufficiently large $N$ such that $g_s N$ is finite for the validity of our analysis. We also note that when $L <  \sqrt{\alpha^{\prime}}$, there does not exist any scale where the gravitational effects play any significant role. Besides, the fact that $L$ in \eqref{8} is independent of the compact volume $V_p$, clearly shows that the concept of a characteristic length scale $L$ is still relevant for the system of branes having infinite extensions in their tangential directions. In our analysis, we consider only the case where the probe particle is farther away from the branes compared to this characteristic length scale.

The paper is organized as follows.
To compare the results of the  motion of a point particle in the transverse space of $p$-branes with that in the  gravitational potential of a point mass in arbitrary $d$-dimensional space-time, we  discuss the results for the latter case first in the next section. In section III, we study the motion of a point particle in the transverse space of $p$-branes. Last section is devoted to a brief conclusion and a discussion of the results.  

\section{General relativity in $d$-dimensions}

In this section we work out the motion of a point particle (both massive and massless) in the gravitational background produced by a point like massive object. Let us note that there has been discussions, in the in the past, on solving geodesic equation for probe particles in Schwarzschild back ground. For example, in a recent paper \cite{hack}, the geodesic equation for a spherically symmetric space time has been exactly solved. The solutions for $d = 4, 5, 7$ are presented in terms of elliptic functions and for $d= 6, 9, 11$ are found in terms of hyperelliptic functions. However, for space time dimensions, $8, 10, \geq 12$ , analytic solutions have not been possible. Since $Dp$-branes are specific to ten dimensional space time and we intend to compare various results of $Dp$-brane background with Schwarzschild background, we  find in this section the leading order solutions  (for motion of light rays) for arbitrary space time dimensions.  Thus,  let us consider a spherically symmetric gravitating object of mass $M$ in $d$-dimensions (where $d\geq 4$) located at $r=0$. The gravitational effects of this object outside is described by the Schwarzschild metric in a $d$-dimensional curved space-time \cite{tangherlini} given by 
\begin{equation}
ds^2 = d \tau^{2} = h(r)dt^2-\frac{1}{h(r)}dr^2 - r^2d\Omega_{d-2}^2,\label{schwarzschild}
\end{equation} 
where we have assumed $c=1$ and 
\begin{equation}
h(r)=\left(1-\frac{2G^{(d)}M}{r^{d - 3}}\right).\label{h(r)}
\end{equation}  
Furthermore, $d\Omega_{d-2}^{2}$ in \eqref{schwarzschild} represents the spherically symmetric angular element and the interesting nontrivial components of the metric tensor follow to be
\begin{equation}
g_{00}=h(r),\qquad g_{11}= g_{rr} = -\frac{1}{h(r)}.
\end{equation} 
This metric describes the static and spherically symmetric vacuum solutions of the $d$-dimensional Einstein's equations. It is easy to see from \eqref{h(r)} that the Schwarzschild radius in this case is given by
\begin{equation}
r_{\sc S}^{(d)} = \left(2G^{(d)}M\right)^{\frac{1}{d-3}}.\label{r_s}
\end{equation}
Let us note that if we compactify $d-4$ spatial dimensions, as we have seen in \eqref{6}, we can relate 
the $d$-dimensional Newton constant to the four dimensional one by the relation
\begin{equation}
G^{(4)} = \frac{G^{(d)}}{V_{d-4}}.\label{compact}
\end{equation}
Therefore, if the radius of compactification is very small, it follows from \eqref{compact} that for a given $G^{(4)},$ the higher dimensional Newton's constant will be smaller, $G^{(d)} \ll G^{(4)}.$ Correspondingly, upon compactification the Schwarzschild radius (for the same mass) will also be smaller in higher dimensions, $r_{\sc S}^{(d)} \ll r_{\sc S}^{(4)}$.

\subsection{Motion of massive particles}

To study the motion of a massive point particle of unit mass in such a background, let us consider the Lagrangian
\begin{equation}
L =  g_{\mu\nu}\dot x^\mu\dot x^\nu,\label{lagrangian}
\end{equation} 
where the metric tensor is given by the Schwarzschild line element \eqref{schwarzschild} and the ``dots" denote derivatives with respect to the proper time (proper length) $\tau$. Since the metric in \eqref{schwarzschild} is static,  $x^{0}=t$ is a cyclic variable and we expect its conjugate momentum to be conserved. In fact, the Euler-Lagrange equation for $t$ gives 
\begin{equation}
\frac{d}{d\tau}\frac{\partial L}{\partial \dot t}=0,\quad\text{or,}\quad  h (r) \dot t = k = \text{constant},\label{tdota}
\end{equation} 
which can be written equivalently as 
\begin{equation}
\dot{t} = \frac{k}{h (r)}.\label{tdot}
\end{equation}

With this brief general introduction, let us study various aspects of a point particle motion in such a background.  Let us  make the following geometric argument to determine the number of coordinates that need to be considered for such a motion in general.  We note that it takes three points to define a two-dimensional plane.  For the first point, we can take the gravitating  object assumed to be at $r=0$.  The second point will be our point test particle of unit mass which feels the gravitational force at  coordinate ${\bf r}(t)$.  We can choose the third point to be the same point mass  only an infinitesimal time later at ${\bf r}(t+dt)$.  These three points define a two-dimensional plane, and the gravitational force will never take us outside of this plane (because the gravitational force is central and pulls the outer point only towards the center radially). This can be understood from the point of view of angular momentum conservation as well. As a result, we can rotate our coordinates so that the metric and the Lagrangian have the simple two spatial dimensional form
\begin{equation}
d\tau^2=h(r)\;dt^2-\frac{1}{h(r)}dr^2-r^2d\phi^2,\quad L = h(r)\;\dot t^2-\frac{1}{h(r)}\dot r^2-r^2\dot \phi^2,\label{co1}
\end{equation} 
where $h (r)$ is given in \eqref{h(r)}. We note from the Lagrangian $L$ in \eqref{co1} that not only is $t$ a cyclic variable (in general for the Schwarzschild metric) as described in \eqref{tdota}, but the angular variable $\phi$ is also a cyclic variable. Therefore, the corresponding conjugate momenta are conserved and the $t,\phi$ equations of motion lead to
\begin{equation}
\dot t= \frac{k}{h(r)} = \frac{k}{1 - \frac{2G^{(d)}M}{r^{d-3}}},\quad \dot \phi=\frac{\ell}{r^2},\label{co2}
\end{equation} 
where $k,\ \ell$ are constants. We recognize $\ell$ to correspond to the angular momentum associated with the particle motion and at this point the two constants appear to be arbitrary. 

\subsubsection{Radial free fall}

Let us begin by studying the motion of a radially (vertically) free falling point particle of unit mass.  In this case $d\phi=0$ (corresponding to the case $\ell=0$), and dividing the line element in \eqref{schwarzschild} by $d\tau^2$ we obtain
\begin{equation}
1= h(r) \dot{t}^{2} - \frac{1}{h(r)}\dot r^2 = \frac{1}{h(r)} (k^{2} - \dot{r}^{2}), \quad{\rm or,}\quad \dot r^2=k^2-h(r),\label{rff1}
\end{equation}
where we have used \eqref{tdot} (or equivalently \eqref{co2}). If we assume that the particle falls initially from rest from a point $r_0$, we have $\dot r\big|_{r_0}=0$,  we determine from \eqref{rff1}
\begin{equation}
k^{2} = h (r_{0}),\quad \Rightarrow\quad  \dot r^2 = h(r_0)-h(r) = 2G^{(d)}M \left(\frac{1}{r^{d-3}} - \frac{1}{r_{0}^{d-3}}\right).\label{rff2}
\end{equation} 
Taking the $\tau$ derivative of \eqref{rff2} we obtain
\begin{align}
 2 \dot{r} \ddot{r} & = - (d-3) \frac{2G^{(d)} M}{r^{d-2}}\ \dot{r},\notag\\
{\rm or,}\quad \ddot{r} & = - \frac{(d-3) G^{(d)} M}{r^{d-2}},\label{rff3}
\end{align}
which is the analogue of Newton's equation in $d$-dimensions. Clearly, in higher dimensions the gravitational force falls off much faster for large $r$ and the attraction is much stronger for smaller $r$ (compared to $4$ dimensions).

From \eqref{rff2} we note that the proper time that a particle will take to come to a radial coordinate $r$ (at $\tau$) is given by
\begin{equation}
\tau = \int\limits_{0}^{\tau}  d\tau^{\prime} = \int\limits_{r_{0}}^{r} \frac{dr^{\prime}}{\dot{r^{\prime}}} = \int\limits_{r}^{r_{0}} dr^{\prime} \left(\frac{(r_{0} r^{\prime})^{d-3}}{2G^{(d)}M (r_{0}^{d-3} - (r')^{d-3})}\right)^{\frac{1}{2}}.\label{rff4}
\end{equation}
Here the sign of $\dot{r}$ (equivalently the limits of the integration) are chosen such that $\tau$ increases as the radial coordinate $r$ decreases. Since the integrand is well behaved, the final integration limit can actually be extended all the way to the Schwarzschild radius (assuming that it lies outside the gravitating mass) and the proper time taken to reach there is easily seen to be finite.

Let us also note that the coordinate speed of the particle is given by
\begin{equation}
v (r) = \frac{dr}{dt} = \frac{\dot{r}}{\dot{t}} = \frac{1 - \frac{2G^{(d)}M}{r^{d-3}}}{\sqrt{1- \frac{2G^{(d)}M}{r_{0}^{d-3}}}}\ \left(\frac{2G^{(d)}M (r_{0}^{d-3} - r^{d-3})}{(r_{0}r)^{d-3}}\right)^{\frac{1}{2}},\label{rff5}
\end{equation}
which shows that the coordinate speed of the particle vanishes at the Schwarzschild radius (as well as at $r=r_{0}$ initially). Since the speed is increasing initially, it must reach a maximum speed at some point before decreasing and the radial coordinate where the coordinate speed is maximum is obtained from
\begin{equation}
\frac{dv(r)}{dr} = 0,\quad \text{or,}\quad r_{\rm max} = \left(\frac{3G^{(d)}Mr_{0}^{d-3}}{2G^{(d)}M + \frac{1}{2} r_{0}^{d-3}}\right)^{\frac{1}{d-3}},\label{rff6}
\end{equation}
where the speed has the value
\begin{equation}
v_{\rm max} = v (r_{\rm max}) = \frac{2}{3\sqrt{3}}\left(1 - \frac{2G^{(d)}M}{r_{0}^{d-3}}\right) = \frac{2}{3\sqrt{3}}\, k^{2}.\label{rff8}
\end{equation}
All these results can be easily seen to reduce to the standard results when $d=4$. From \eqref{rff6} and \eqref{rff8} we see that for $r_{0}\gg r_{\sc S}$ ($c=1$)
\begin{equation}
r_{\rm max} = \left(6G^{(d)}M\right)^{\frac{1}{d-3}} = \left(3\right)^{\frac{1}{d-3}}\, r_{\sc S},\quad v_{\rm max} = v(r_{\rm max}) = \frac{2}{3\sqrt{3}},\label{rff9}
\end{equation}
This shows that, in this case, the maximum speed is reached very close to the Schwarzschild radius. In fact, $1.7 r_{\sc S}^{(d)}\leq r_{\rm max}\leq 3r_{\sc S}^{(d)}$ as $10\leq d \leq 4$ and for $r_{0} \gg r_{\sc S}$ the maximum speed achieved by the test particle is independent of the number of space-time dimensions.

\subsubsection{Circular orbit}

We will now discuss more general solutions of the particle motion. First, let us consider circular motion of the particle. To begin with let us note that classical Newtonian gravity does not permit bound circular orbits in higher dimensions $d > 4$. To see this, let us recall that the Lagrangian for the Newtonian motion in $d$ dimensions is given by (the test particle has unit mass)
\begin{equation}
L = \frac{1}{2} \left(\left(\frac{dr_{\sc N}}{dt}\right)^{2} + r_{\sc N}^{2} \left(\frac{d\phi}{dt}\right)^{2}\right) + \frac{G^{(d)}M}{r_{\sc N}^{d-3}},\label{co3a}
\end{equation}
and the energy associated with the particle motion is obtained to be
\begin{equation}
E = \frac{1}{2}\left(\left(\frac{dr_{\sc N}}{dt}\right)^{2} + r_{\sc N}^{2}\left(\frac{d\phi}{dt}\right)^{2}\right) - \frac{G^{(d)}M}{r_{\sc N}^{d-3}} < 0,\label{co3b}
\end{equation}
for bound motion. The $\phi$-equation leads to the conservation law
\begin{equation}
\frac{d\phi}{dt} = \frac{\ell}{r_{\sc N}^{2}},\label{co3c}
\end{equation}
and the radial equation for circular motion ($\frac{dr_{\sc N}}{dt} = 0 = \frac{d^{2}r_{\sc N}}{dt^{2}}$) leads to
\begin{equation}
r_{\sc N}\left(\frac{d\phi}{dt}\right)^{2} = \frac{(d-3)G^{(d)}M}{r_{\sc N}^{d-2}}.\label{co3d}
\end{equation}
Equation \eqref{co3d} simply expresses the fact that, for circular motion, the centrifugal force must balance the gravitational attraction. Using \eqref{co3c} we can determine the radius of a possible circular orbit $r_{\sc N}$ from \eqref{co3d} to satisfy
\begin{equation}
r_{\sc N}^{d-5} = \frac{(d-3)G^{(d)} M}{\ell^{2}}.\label{co3e}
\end{equation}
Clearly, there are two distinct cases to consider. For $d=5$, \eqref{co3e} leads to the constraint
\begin{equation}
\ell^{2} = 2G^{(5)} M,\label{co3e'}
\end{equation}
for which we determine from \eqref{co3b} that
\begin{equation}
E = 0.
\end{equation}
For $d > 5$, \eqref{co3e} determines 
\begin{equation}
r_{\sc N} = \left(\frac{(d-3) G^{(d)} M}{\ell^{2}}\right)^{\frac{1}{d-5}},\label{co3e''}
\end{equation}
which leads to the energy associated with motion to be
\begin{equation}
E = \frac{\ell^{2}}{2r_{\sc N}^{2}}\, \frac{d-5}{d-3} > 0,\quad d > 5.
\end{equation}
In either case we note that this does not correspond to bound motion for which $E < 0$. 

This can be understood more physically as follows. Let us define an effective potential associated with the particle motion as (see \eqref{co3b})
\begin{equation}
V_{\rm eff} (r_{\sc N}) = \frac{\ell^{2}}{2r_{\sc N}^{2}} - \frac{G^{(d)} M}{r_{\sc N}^{d-3}},\label{co3f}
\end{equation}
where the first term represents the centrifugal potential which is independent of the space-time dimensions while the second term representing gravitational attraction does depend on $d$. When $d=4$ the positive centrifugal potential dominates the gravitational attraction for small values of $r_{\sc N}$ while for large $r_{\sc N}$ it is the other way around. This leads to a stable minimum of the potential at some negative value (for \eqref{co3b} to be true with $r_{\sc N}$ constant) for a finite $r_{\sc N}$ where the centrifugal force can equal the gravitational attraction. For $d=5$, both the terms in the effective potential have the same $\frac{1}{r_{\sc N}^{2}}$ behavior and if $\ell^{2} = 2 G^{(d)}M$, the effective potential vanishes and is incompatible with \eqref{co3b}. For $d > 5$, the gravitational attraction dominates the positive centrifugal attraction for small values of $r_{\sc N}$ while it is the other way around for large $r_{\sc N}$. As a result, the extremum of the effective potential in this case is an unstable maximum and occurs at a positive value of the potential (centrifugal potential is positive) which is inconsistent with \eqref{co3b}. Alternatively we note that the extremum of the effective potential is derived from
\begin{equation}
V_{\rm eff}' (r_{\sc N}) = - \frac{1}{r_{\sc N}^{3}}\left(\ell^{2} - \frac{(d-3) G^{(d)} M}{r_{\sc N}^{d-5}}\right) = 0,\label{co3g}
\end{equation}
which determines 
\begin{equation}
d=5:\  \ell^{2} - 2G^{(5)} M = 0,\quad d > 5:\ r_{\sc N} = \left(\frac{(d-3) G^{(d)} M}{\ell^{2}}\right)^{\frac{1}{d-5}}.\label{co3g'}
\end{equation}
This can be compared with the earlier results in \eqref{co3e'} and \eqref{co3e''}. Furthermore, the second derivative of the potential at the extremum gives
\begin{equation}
d=5:\ V_{\rm eff}'' (r_{\sc N}) = 0,\quad d > 5:\ V_{\rm eff}'' (r_{\sc N}) = \frac{\ell^{2}}{r_{\sc N}^{4}}\left(5 - d\right) < 0,\label{co3h}
\end{equation}
implying that the extremum is an unstable maximum (for $d > 5$) which cannot support bound state motion.

On the other hand, inclusion of relativistic effects does allow for stable circular motion. For example, the $r$-equation following from \eqref{co1} leads to ($\dot{r} = \ddot{r} = 0$)
\begin{equation}
r \dot{\phi}^{2} =  \frac{(d-3) G^{(d)}M}{r^{d-2}}\, \dot{t}^{2},\label{co3}
\end{equation}
and this equation (compare with \eqref{co3d}) clearly shows that the two constants defined in \eqref{co2} are related. Explicitly we have
\begin{equation}
\ell^{2} = \frac{(d-3) G^{(d)}M}{r^{d-5}}\,\frac{k^{2}}{\left(1 - \frac{2G^{(d)}M}{r^{d-3}}\right)^{2}}.\label{kell}
\end{equation} 
Using  \eqref{co3} in the relation following from the line element for circular motion, namely,
\begin{equation}
1 = h (r) \dot{t}^{2} - r^{2} \dot{\phi}^{2},\label{co3''}
\end{equation}
as well as the relations \eqref{co2} shows that \eqref{co3} leads to
\begin{equation}
\left(1 - \frac{(d-1) G^{(d)} M}{r^{d-3}}\right)\frac{k^{2}}{\left(1- \frac{2G^{(d)} M}{r^{d-3}}\right)^{2}} = 1.\label{co3'''}
\end{equation}
This relation is consistent only for 
\begin{equation}
r > \left(\frac{(d-1)}{2}\right)^{\frac{1}{d-3}} r_{\sc S}^{(d)},\label{co10}
\end{equation}
showing the possibility of circular motion. In fact, \eqref{co3'''} determines the value of the constant (for a fixed $r$) to be
\begin{equation}
k = \frac{\left(1 - \frac{2G^{(d)}M}{r^{d-3}}\right)}{\sqrt{1 - \frac{(d-1)G^{(d)}M}{r^{d-3}}}}.\label{co8}
\end{equation}

To understand circular motion better, let us use \eqref{co3} to determine
\begin{align}
\frac{d\phi}{dt} & = \sqrt{\frac{(d-3)G^{(d)}M}{r^{d-1}}},\notag\\
{\rm or,}\quad \int dt & = \sqrt{\frac{r^{d-1}}{(d-3)G^{(d)}M}}\ \int d\phi,\label{co4}
\end{align}
where we have assumed $r$ is a constant for circular motion. Equation \eqref{co4} determines the (coordinate) period of the orbit of such a motion to be
\begin{equation}
T = \Delta t = 2\pi\left(\frac{r^{d-1}}{(d-3) G^{(d)}M}\right)^{\frac{1}{2}}.\label{co5}
\end{equation}
This can be thought of as the generalization of Kepler's law to $d$-dimensions, namely, the square of the period of the orbit is proportional to the radius of the orbit raised to $(d-1)$ power which corresponds to the number of (transverse) spatial dimensions in $d$ space-time dimensions and this reduces to the standard result when $d=4$. (One should keep in mind though that these are only coordinate parameters.) 

To determine the proper period associated with such a motion, we note from the first relation in \eqref{co2} that we can write
\begin{equation}
\Delta \tau = \left(1 - \frac{2G^{(d)}M}{r^{d-3}}\right) \frac{\Delta t}{k},\label{co6}
\end{equation}
and using the value of $k$ from \eqref{co8} leads the proper period \eqref{co6} to correspond to
\begin{equation}
\Delta \tau = \left(1 - \frac{(d-1) G^{(d)}M}{r^{d-3}}\right)^{\frac{1}{2}} \Delta t.\label{co9}
\end{equation}
We note that the proper period vanishes for
\begin{equation}
r = \left(\frac{(d-1)}{2}\right)^{\frac{1}{d-3}} r_{\sc S}^{(d)},\label{co11}
\end{equation}
which would correspond to the circular orbit for the photon in $d$-dimensions (which has a vanishing proper time) and this coincides with the well known result when $d=4$.

\subsubsection{General motion}\label{sgm}

If we do not assume the particle to be moving in a circular orbit, the radius will not be a constant so that  $\dot{r} \neq 0, \ddot{r} \neq 0$. In this case, the $r$ equation is different from \eqref{co3} and, consequently, the two constants $k,\ \ell$ defined by \eqref{co2} are not directly related. In fact, the $r$ equation can be easily obtained from the line element in \eqref{co1} which leads to
\begin{align}
\left(\frac{dr}{d\phi}\right)^{2} & = - \frac{\left(1 - \frac{2G^{(d)}M}{r^{d-3}}\right)}{\dot{\phi}^{2}} + \left(1 - \frac{2G^{(d)}M}{r^{d-3}}\right)^{2}\frac{\dot{t}^{2}}{\dot{\phi}^{2}} - r^{2}\left(1 - \frac{2G^{(d)}M}{r^{d-3}}\right)\notag\\
& = \frac{r^{4}}{\ell^{2}}\left((k^{2}-1) + \frac{2G^{(d)}M}{r^{d-3}}\right) - r^{2}\left(1 - \frac{2G^{(d)}M}{r^{d-3}}\right),\label{gm1}
\end{align}
where we have used the two relations in \eqref{co2} and $\ell$ is assumed to be nonzero. ($\ell =0$ corresponds to the case of radial fall which we have discussed earlier. For $\ell$ and $k$ defined by \eqref{kell} the right hand side of \eqref{gm1} vanishes leading to a constant $r$, namely, a circular motion.) Defining $u = \frac{1}{r}$, this equation leads to
\begin{equation}
\left(\frac{du}{d\phi}\right)^{2} - \frac{2G^{(d)} M u^{d-3}}{\ell^{2}} + u^{2} \left(1- 2G^{(d)} M u^{d-3}\right) = \frac{(k^{2}-1)}{\ell^{2}}.\label{gm2}
\end{equation}
Upon differentiation, this gives two possible equations
\begin{equation}
\frac{du}{d\phi} = 0,\quad \text{or,}\quad \frac{d^{2}u}{d\phi^{2}} = \frac{(d-3) G^{(d)}M}{\ell^{2}}\,u^{d-4} - u + (d-1) G^{(d)}M u^{d-2}.\label{gm3}
\end{equation}
The first equation in \eqref{gm3} corresponds to the circular motion which we have already discussed. The second equation, on the other hand, describes the more general motion of a massive particle. In $d=4$, we know that the first two terms on the right hand side lead to elliptic orbits whereas the last term corresponds to relativistic correction to these orbits resulting in the perihelion advance. Correspondingly, the equation
\begin{equation} 
\frac{d^{2}u_{\sc N}}{d\phi^{2}} = - u_{\sc N} + \frac{(d-3) G^{(d)}M}{\ell^{2}}\,u_{\sc N}^{d-4},\label{gm4}
\end{equation}
can be thought of as representing the classical Kepler's equation in $d$-dimensions for a massive particle moving in the gravitational potential \eqref{4}. The last term in \eqref{gm3} would then represent the relativistic correction to this equation. 

The Newtonian relation for bound motion analogous to \eqref{gm2} (or \eqref{co3b}) has the form
\begin{equation}
\left(\frac{du_{\sc N}}{d\phi}\right)^{2} + u_{\sc N}^{2} - \frac{2G^{(d)} M u_{\sc N}^{d-3}}{\ell^{2}} = \frac{2E}{\ell^{2}} < 0.\label{gm5}
\end{equation}
Equation \eqref{gm4} is solved in closed form for $d = 4, 5, 6, 7, 9, 11$ and could not be solved for other values of $d$ \cite{hack}.  However, the analysis of the stability of any bound orbits can be done easily from \eqref{gm5} (see \eqref{co3g}-\eqref{co3h}). Since $d=4$ is well studied, let us restrict to $d \geq 5$. In \eqref{gm5} we can identify
\begin{equation}
V_{\rm eff} (u_{\sc N}) =  u_{\sc N}^{2} - \frac{2G^{(d)} M u_{\sc N}^{d-3}}{\ell^{2}},\label{gm6}
\end{equation}
which coincides with \eqref{co3f} (up to a multiplicative constant) and, therefore, the stability analysis of the motion goes exactly as in \eqref{co3g}-\eqref{co3h} showing that no bound motion is possible. Intuitively this can be understood from the fact that for $r\gg r_{\sc S}$ (meaningful for physical motion) the attractive gravitational potential falls off much faster for $d > 5$ than the repulsive centrifugal potential leading to unbounded motion.

On the other hand, including the relativistic correction, we can define the effective potential from \eqref{gm2} to correspond to
\begin{equation}
V_{\rm eff} (u) =  - \frac{2G^{(d)} M u^{d-3}}{\ell^{2}} + u^{2} \left(1- 2G^{(d)} M u^{d-3}\right).\label{gm7}
\end{equation}
The extremum of this potential is obtained from
\begin{equation}
V_{\rm eff}' (u) = 2u \left(1 - \frac{(d-3) G^{(d)} M}{\ell^{2}}\,u^{d-5} - (d-1) G^{(d)} M u^{d-3}\right) = 0.\label{gm8}
\end{equation}
This equation has one vanishing root and only one positive root $u_{0}$ (which is physical) for $d\geq 5$ (this follows easily from Descartes' rule of signs). The second derivative of the potential at the positive root leads to
\begin{equation}
V_{\rm eff}'' (u_{0}) = - 2 \left((d-5) + 2 (d-1) G^{(d)} M u_{0}^{d-3}\right) < 0,\quad d\geq 5,\label{gm9}
\end{equation}
leading to instability of motion. Once again, we see that the relativistic correction in \eqref{gm7} only adds an attractive potential which falls off even faster than the classical gravitational potential and thereby does not help in stabilizing the motion. Thus, unlike the circular motion, including relativistic correction does not stabilize the general motion and this was the main thrust of the analysis in \cite{tangherlini}.

\subsection{Motion of light rays}

Finally, we discuss the motion of photons in the higher dimensional gravitational field.  As we have already seen in \eqref{co11}, photons can move in circular orbits of radius (as we have emphasized earlier, our analysis is for $d\geq 4$)
\begin{equation}
r = \left(\frac{(d-1)}{2}\right)^{\frac{1}{d-3}} r_{s}^{(d)} = \left((d-1)G^{(d)} M\right)^{\frac{1}{d-3}}.\label{lr1}
\end{equation} 
Normally, such a radius lies inside a star and, therefore, is not physical. There is no other bound motion possible for the photons. Therefore, in this section we will discuss the phenomenon of bending of light in the vicinity of a gravitating star in $d$ dimensions.

Since photons are massless particles, the proper time associated with them vanishes and we have $d\tau^2=0$. Consequently, we cannot label the photon trajectory with $\tau$ and have to introduce an affine parameter $\lambda$ for this purpose. Equations \eqref{co2} continue to hold
\begin{equation}
\dot t=\frac{k}{h(r)} = \frac{k}{1 - \frac{2G^{(d)}M}{r^{d-3}}},\quad \dot \phi=\frac{\ell}{r^2},\label{lr2}
\end{equation} 
where the dots now refer to derivatives with respect to $\lambda$. Only the radial equation modifies which can be obtained easily from the relation for the line element. Since $d\tau^{2}=0$ it follows that
\begin{equation}
\left(\frac{dr}{d\phi}\right)^{2} + \left(1 - \frac{2G^{(d)}M}{r^{d-3}}\right) r^{2} - \frac{k^{2}r^{4}}{\ell^{2}} = 0.\label{lr3}
\end{equation}
In terms of the variable $u = \frac{1}{r}$, this can be written as
\begin{equation}
\left(\frac{du}{d\phi}\right)^{2} + (1 - 2G^{(d)} M u^{d-3}) u^{2} = \frac{k^{2}}{\ell^{2}},\label{lr4}
\end{equation}
which can be compared with \eqref{gm2}. Taking derivative of this equation with respect to $u$ we obtain
\begin{equation}
\frac{du}{d\phi} \left(\frac{d^{2}u}{d\phi^{2}} + u - (d-1) G^{(d)}M u^{d-2}\right) = 0,\label{lr5}
\end{equation}
which leads to two possibilities
\begin{equation}
\frac{du}{d\phi} = 0,\quad \frac{d^{2}u}{d\phi^{2}} + u - (d-1) G^{(d)}M u^{d-2} = 0.\label{lr6}
\end{equation}
The first of these describes the circular orbit that we have already discussed. Therefore, let us analyze the second equation
\begin{equation}
\frac{d^{2}u}{d\phi^{2}} + u = (d-1) G^{(d)}M u^{d-2} = \epsilon u^{d-2}.\label{lr6a}
\end{equation}
where we have identified the relativistic correction as 
\begin{equation}
(d-1) G^{(d)}M u^{d-2} = \epsilon u^{d-2},\label{lr7}
\end{equation}
and recognize that it represents a very small correction compared to the centrifugal force for $r \gg r_{s}^{(d)}$. 

The solution to equation \eqref{lr6a} without the relativistic correction is that of a straight line which can be written as
\begin{equation}
u (\phi) = u_{\rm min} \sin\phi,\label{lr8}
\end{equation}
where we have chosen the closest distance of approach of the photon to the star $r_{\rm min} = \frac{1}{u_{\rm min}}$ to occur at $\phi = \frac{\pi}{2}$. Using this we can write down the leading solution (in powers of $\epsilon$) to \eqref{lr6a} of the form
\begin{equation}
u (\phi) = u_{\rm min} \sin \phi + \epsilon u_{1} (\phi).\label{lr9}
\end{equation}
Substituting this back into \eqref{lr6a} we obtain to linear order in $\epsilon$
\begin{equation}
\frac{d^{2}u_{1}}{d\phi^{2}} + u_{1} = u_{\rm min}^{d-2} \sin^{d-2} \phi.\label{lr10}
\end{equation}
The solution of this equation in even dimensions can be written in the form
\begin{equation}
u_{1} (\phi) = \sum\limits_{n=1}^{\frac{d}{2}} \alpha_{n} \sin^{d-2n} \phi,\label{lr11}
\end{equation}
with the constant coefficients given by
\begin{equation}
\alpha_{1} = - \frac{u_{\rm min}^{d-2}}{(d-1)(d-3)},\quad \alpha_{n+1} = \frac{(d-2n)}{(d-2n-3)}\,\alpha_{n},\quad n=1,2,\cdots, \frac{d}{2}-1.\label{lr12}
\end{equation}
Therefore, the complete solution to the leading order in $\epsilon$ can be written in even dimensions as
\begin{equation}
u (\phi) = u_{\rm min} \sin \phi + (d-1) G^{(d)} M \sum\limits_{n=1}^{\frac{d}{2}} \alpha_{n} \sin^{d-2n} \phi.\label{lr13}
\end{equation}
This is easily seen to reduce to the well studied results when $d=4$. 

Let us next note that in the absence the relativistic correction, the solution \eqref{lr8} shows that the light ray would come in from infinity at an angle $\phi =0$ and go out to infinity at $\phi=\pi$. In the presence of the relativistic correction we expect the light ray to come in from infinity at an angle $\phi = -\delta$ and go out to infinity at $\phi = \pi + \delta$ (by symmetry) where we expect the angle $\delta$ to be small. Using this in \eqref{lr13} we obtain
\begin{equation}
0 = u_{\rm min} \sin (-\delta) + (d-1) G^{(d)}M \sum\limits_{n=1}^{\frac{d}{2}} \alpha_{n} \sin^{d-2n} (-\delta).\label{lr14}
\end{equation}
Recognizing that $\delta$ is a small angle, we determine 
\begin{equation}
\delta \approx \frac{1}{u_{\rm min}}\, (d-1) G^{(d)} M \alpha_{\frac{d}{2}} = - \frac{G^{(d)} M u_{\rm min}^{d-3}}{(d-3)} \prod\limits_{n=1}^{\frac{d}{2}-1} \left(1 + \frac{3}{(d-2n-3)}\right).\label{lr15}
\end{equation}
The bending of the light path, therefore, is given by
\begin{equation}
\Delta = 2\delta \approx - \frac{2G^{(d)} M u_{\rm min}^{d-3}}{(d-3)} \prod\limits_{n=1}^{\frac{d}{2}-1} \left(1 + \frac{3}{(d-2n-3)}\right).\label{lr16}
\end{equation}
We note here that the overall negative sign reflects the fact that the product on the right hand side is negative.

In odd dimensions (recall that $d\geq 5$), the solution of \eqref{lr10} can be obtained in a similar fashion and has the form
\begin{equation}
u_{1} (\phi) = \beta \phi \cos \phi + \sum\limits_{n=1}^{[\frac{d}{2}] -1} \alpha_{n} \sin^{d-2n} \phi,\label{lr17}
\end{equation}
where $[\frac{d}{2}]$ denotes the maximum integer part of $\frac{d}{2}$ and the coefficients satisfy
\begin{equation}
\alpha_{1} = - \frac{u_{\rm min}^{d-2}}{(d-1)(d-3)},\quad \alpha_{n+1} = \frac{(d-2n)}{(d-2n-3)}\,\alpha_{n},\quad \beta = 3 \alpha_{[\frac{d}{2}]-1},\quad n=1,2,\cdots , [\frac{d}{2}]-2.\label{lr18}
\end{equation}
In this case, we can also carry out the calculation for the photon deflection in an analogous manner. The equations force that the light ray comes in from infinity at $\phi =0$ and goes out at $\phi = \pi + \delta$ and the leading value of the total bending is determined to be
\begin{equation}
\Delta = \delta \approx \frac{3\pi G^{(d)}M u_{\rm min}^{d-3}}{(d-3)} \prod\limits_{n=1}^{[\frac{d}{2}]-2} \left(1 + \frac{3}{(d-2n-3)}\right).\label{lr19}
\end{equation}

\section{Motion in transverse space of D$p$-branes}
    
In the previous section we systematically analyzed the motion of a point particle in a Schwarzschild background in higher dimensions and discussed various consequences associated with it. In this section we will carry out the corresponding analysis for the motion of a point particle in the transverse space of a stack of D$p$-branes in ten dimensions ($d=10$). The gravitational field produced by the stack of branes is given in \eqref{2} and, as before, we can show that the motion of the test particle will be planar so that the invariant length of the probe particle takes the form ($c=1$)
\begin{equation}
ds^2 = d\tau^{2} = H^{-\frac{1}{2}} (r) dt^2-H^{\frac{1}{2}} (r)\left(dr^2+r^2d\phi^2\right),\label{b1}
\end{equation}
where the harmonic function $H (r)$ is defined in \eqref{3}. The nontrivial components of the metric tensor are obtained from \eqref{b1} to correspond to
\begin{equation}
g_{00} = H^{-\frac{1}{2}} (r),\quad g_{11} = g_{rr} = - H^{\frac{1}{2}} (r),\quad g_{22}=g_{\phi\phi} = - r^{2} H^{\frac{1}{2}}.\label{b1a}
\end{equation}

\subsection{Motion of massive particles} 

Let us assume that the test particle has a unit mass so that the dynamics of the particle is described by the Lagrangian \eqref{lagrangian}
\begin{equation}
L = g_{\mu\nu}\dot x^\mu\dot x^\nu,\label{lagrangian1}
\end{equation}
where $\dot x^{\mu} = \frac{dx^{\mu}}{d\tau}$ and the components of the metric tensor $g_{\mu\nu}$ are given in  \eqref{b1a}. As before we note that $t$ and $\phi$ are cyclic variables so that the corresponding conjugate momenta are conserved. From the $t, \phi$ equations, it follows that (compare with \eqref{co2})
\begin{equation}
\dot t = \frac{k}{g_{00}} = kH^{\frac{1}{2}} (r),\quad \dot\phi = -\frac{\ell}{g_{\phi\phi}}=\frac{\ell}{r^2}\, H^{-\frac{1}{2}} (r),\label{b2}
\end{equation}
where $k$ and $\ell$ are constants. With this let us now analyze the different types of motion the (massive) point particle can have in such a gravitational background.

\subsubsection{Radial free fall}

Let us start with the radial free fall of the point particle in this gravitational background. In this case we have $\ell=0$ since only the radial coordinate changes ($d\phi=0$) and from \eqref{b1} we obtain (using \eqref{b2}) 
\begin{equation}
1=H^{\frac{1}{2}} (r) (k^2-\dot r^2),\quad \Rightarrow\quad \dot r^2=k^2-H^{-\frac{1}{2}} (r).\label{brf1}
\end{equation}
If we assume that the particle falls from rest from a point $r_0$, we have $\dot r|_{r_0}=0$. As a result, we see from \eqref{brf1} that 
\begin{equation}
k^2=H^{-\frac{1}{2}} (r_{0}),\quad\dot r^2=H^{-\frac{1}{2}} (r_{0}) - H^{-\frac{1}{2}} (r).\label{brf2}
\end{equation}
Taking the $\tau$ derivative of the second equation in \eqref{brf2} we obtain
\begin{align}
2 \dot{r} \ddot{r} & = - \frac{7-p}{2}\,H^{-\frac{3}{2}} (r)\, \frac{Q_{p}}{r^{8-p}}\,\dot{r},\notag\\
{\rm or,}\quad \ddot{r} & =  - \frac{7-p}{4}\,H^{-\frac{3}{2}} (r)\, \frac{Q_{p}}{r^{8-p}},\label{brf3}
\end{align}
which can be thought of as the analogue of Newton's equation in such a background. This equation can be compared with \eqref{rff3} and it is clear that as $p$ increases, the gravitational force due to the branes falls off more slowly for large $r$ than that due to the Schwarzschild background (for $d=10$). This is a reflection of the fact that the number of transverse spatial dimensions decreases for higher dimensional  extended objects. For short distances (see \eqref{3a}), the gravitational force, in fact, decreases for $p < 5$, becomes a constant for $p = 5$ and grows slowly for $p = 6$. This should be contrasted with \eqref{rff3} where the gravitational force increases quite strongly as $r$ becomes small. 

The coordinate velocity can be obtained from \eqref{brf2} and is given by
\begin{equation}
v (r) = \frac{dr}{dt}=\frac{\dot r}{\dot t} = -H^{\frac{1}{4}} (r_{0}) H^{-\frac{1}{2}} (r) \left(H^{-\frac{1}{2}} (r_0) - H^{-\frac{1}{2}} (r)\right)^{\frac{1}{2}} = - H^{-\frac{1}{2}} (r_{0}) G (r) \left(1 - G (r)\right)^{\frac{1}{2}},\label{brf4}
\end{equation}
where we have defined
\begin{equation}
G (r) = H^{\frac{1}{2}} (r_{0}) H^{-\frac{1}{2}} (r) = \left(\frac{H (r_{0})}{H (r)}\right)^{\frac{1}{2}}.\label{brf5}
\end{equation}
It now follows that
\begin{equation}
\frac{d v(r)}{dr} = - (2 - 3 G (r))\left(\frac{1}{4H (r_{0}) (1 - G (r))}\right)^{\frac{1}{2}}\, \frac{dG (r)}{dr} = 0,\label{brf6}
\end{equation}
for $G (r_{\rm max}) = \frac{2}{3}$. Since the radial coordinate is decreasing with time, $H (r) > H (r_{0})$ and such a solution is possible. The coordinate where the velocity will be a maximum is given by
\begin{equation}
r_{\rm max}=\left(\frac{5}{4Q_p}+\frac{9}{4r_0^{7-p}}\right)^{\frac{1}{p-7}} \xrightarrow{\text{large}\ r_{0}} \left(\frac{4Q_{p}}{5}\right)^{\frac{1}{7-p}},\label{brf7}
\end{equation}
and the magnitude of the (maximum) velocity at that point is given by
\begin{equation}
|v (r_{\rm max})| = H^{-\frac{1}{2}} (r_{0}) G (r_{\rm max}) (1 - G (r_{\rm max}))^{\frac{1}{2}} = \frac{2}{3\sqrt{3}} H^{-\frac{1}{2}} (r_{0}) = \frac{2}{3\sqrt{3}}\, k^{2},\label{brf8}
\end{equation}
which can be compared with \eqref{rff8}. However, note that for a given $r_{0}$, the constant $k^{2}$ is different for the Schwarzschild and the brane backgrounds. Nonetheless, it is worth emphasizing that in either case, for large enough $r_{0}$, the maximum speed a test particle can achieve in a radial fall is given by (see also \eqref{rff9} and note that $c=1$)
\begin{equation}
v (r_{\rm max}) = \frac{2}{3\sqrt{3}}.\label{brf9}
\end{equation}
The other significant point to note from \eqref{brf4} is that unlike the Schwarzschild background where the point particle has vanishing coordinate velocity at the Schwarzschild radius (in addition to the initial point, see \eqref{rff5} and discussion there), here the coordinate velocity does not seem to vanish (besides the initial point) except at the singular point $r=0$.

\subsubsection{Circular orbit}

Let us next consider a slightly more general motion of the massive test particle (of unit mass), namely, the motion in a circular orbit around the gravitational source. The line element \eqref{b1}, in general, leads to
\begin{equation}
1 = H^{-\frac{1}{2}} (r) \dot{t}^{2} - H^{\frac{1}{2}} (r) \left(\dot{r}^{2} + r^{2} \dot{\phi}^{2}\right),\label{bco1}
\end{equation}
where dots denote derivatives with respect to the proper time $\tau$. If we are interested in a circular orbit, then $\dot{r} = 0 = \ddot{r}$ and \eqref{bco1} leads to
\begin{equation} 
1 = H^{-\frac{1}{2}} (r) \dot{t}^{2} - r^{2} H^{\frac{1}{2}} (r) \dot{\phi}^{2}.\label{bco2}
\end{equation}
On the other hand, the radial equation from \eqref{lagrangian} (or \eqref{lagrangian1}) leads to
\begin{equation}
- \frac{\partial H^{-\frac{1}{2}} (r)}{\partial r}\, \dot{t}^{2} + \frac{\partial (r^{2} H^{\frac{1}{2}} (r))}{\partial r}\, \dot{\phi}^{2} = 0.\label{bco3}
\end{equation}
Equations \eqref{bco2} and \eqref{bco3} show that the two constants $k,\ell$ are not independent in this case. Explicitly, using \eqref{b2} in \eqref{bco2} we obtain
\begin{equation}
\ell^{2} = r^{2} H^{\frac{1}{2}} (r) \left(k^{2} H^{\frac{1}{2}} (r) - 1\right),\label{kell1}
\end{equation}
which can be compared with \eqref{kell}.  Equation \eqref{bco3} allows us to determine
\begin{equation}
H^{\frac{1}{2}} (r) r^{2} \dot{\phi}^{2} = - \frac{r H^{-\frac{1}{2}} (r) H^{\prime} (r)}{4H (r) + r H^{\prime} (r)}\,\dot{t}^{2},\label{bco4}
\end{equation}
where a prime denotes a derivative with respect to $r$. Substituting this into \eqref{bco2} and using the definition of $k$ from \eqref{b2} we determine 
\begin{equation}
k = \sqrt{\frac{1}{2H^{\frac{1}{2}} (r)}}\left(1 + \frac{2 H (r)}{2 H (r) + r H^{\prime} (r)}\right)^{\frac{1}{2}}.\label{bco5}
\end{equation}

We note from equation \eqref{bco4} that we can write
\begin{equation}
\left(\frac{dt}{d\phi}\right)^{2} = \left(4 r^{7-p} + (p-3)Q_{p}\right) \frac{r^{2} H (r)}{(7-p)Q_{p}}.\label{bco6}
\end{equation}
It is clear from \eqref{bco6} that the right hand side is positive for any value of $r$ for $p \geq 3$ (recall that we are considering $0\leq p \leq 6$, see discussion following \eqref{3a}) and, therefore, circular orbits are possible in general in these cases. When $p < 3$, the right hand side is positive only for 
\begin{equation}
r > \left(\frac{(3-p)Q_{p}}{4}\right)^{\frac{1}{7-p}},\label{bco7}
\end{equation}
and circular orbits are allowed only for these values of the radial coordinate. The period of the orbit can now be determined from \eqref{bco6} to be
\begin{equation}
T = \Delta t  = 2\pi \left(\frac{1}{(7-p)Q_{p} r^{5-p}} \left(4 r^{7-p} + (p-3)Q_{p}\right)\left(r^{7-p} + Q_{p}\right)\right)^{\frac{1}{2}},\label{bco8}
\end{equation}
which represents the generalization of Kepler's law for such a gravitational background. It does not have a particularly simple form. However, for large values of $r$ we note that
\begin{equation}
T \xrightarrow{\text{large}\ r} r^{\frac{9-p}{2}},\label{bco9}
\end{equation}
which can be compared with \eqref{co5} (with $d=10$) and shows that the square of the time period has a smaller power growth with radius as the brane dimension increases (compared with the Schwarzschild background). In general, for any background and for large radial coordinates it seems that Kepler's law generalizes as the square of the period is proportional to $r^{\alpha}$ where $\alpha$ denotes the number of transverse spatial dimensions associated with the gravitational source. Finally, we note that the proper period associated with the circular orbit can be determined to have the form
\begin{equation}
\Delta \tau = H^{-\frac{1}{2}} (r) \frac{\Delta t}{k} = \sqrt{2} H^{-\frac{1}{4}}\left(1 + \frac{2 H (r)}{2 H(r) - \frac{(7-p)Q_{p}}{r^{7-p}}}\right)^{-\frac{1}{2}}\, \Delta t.\label{bco10}
\end{equation}
It is interesting to note that the prefactor multiplying $\Delta t$ in \eqref{bco10} vanishes for $p<3$ precisely for (see also \eqref{bco7})
\begin{equation}
r = \left(\frac{(3-p)Q_{p}}{4}\right)^{\frac{1}{7-p}},\label{bco11}
\end{equation}
and would correspond to the radius for the circular orbit of a photon. However, for $p\geq 3$, the prefactor does not vanish and correspondingly, the photon cannot have a circular orbit for such D$p$-branes.

\subsubsection{Perihelion advance}

Let us next analyze the general (noncircular orbial) bound motion of a massive point particle (of unit mass) in the background of a stack of D$p$ branes. From the equation for the line element \eqref{bco1} we have
\begin{equation}
\left(\frac{dr}{d\phi}\right)^{2} + r^{2} - H^{-1} (r) \frac{\dot{t}^{2}}{\dot{\phi}^{2}} + \frac{H^{-\frac{1}{2}}}{\dot{\phi}^{2}} = 0.\label{bpa1}
\end{equation}
Using the relations in \eqref{b2}, we can write \eqref{bpa1} as
\begin{equation}
\left(\frac{dr}{d\phi}\right)^{2} + r^{2} - \frac{r^{4}}{\ell^{2}}\, H^{\frac{1}{2}} (r) \left(k^{2} H^{\frac{1}{2}} (r) - 1\right) = 0,\label{bpa2}
\end{equation}
and this shows that for $k, \ell$ satisfying \eqref{kell1}, this leads to a circular motion which we are not considering. As before, let us introduce the variable $u = \frac{1}{r}$ so that \eqref{bpa2} can be written as
\begin{equation}
\left(\frac{du}{d\phi}\right)^{2} + u^{2} - \frac{k^{2}}{\ell^{2}} Q_{p} u^{7-p} + \frac{1}{\ell^{2}} \left(1 + Q_{p} u^{7-p}\right)^{\frac{1}{2}} = \frac{k^{2}}{\ell^{2}}.\label{bpa3}
\end{equation}
This can be the starting point of our analysis for the stability of motion as we have done in subsection \ref{sgm}. However, because of the square root, a general analysis of this relation becomes quite complicated. On the other hand, we note that  for physical motion we expect $Q_{p} u^{7-p} \ll 1$  and, in this case, we can approximate
\begin{equation}
\left(1 + Q_{p}u^{7-p}\right)^{\frac{1}{2}} = 1 + \frac{Q_{p}}{2} u^{7-p} - \frac{Q_{p}^{2}}{8} u^{14-2p} + \frac{Q_{p}^{3}}{16} u^{21 - 3p} + O (Q_{p}^{4}),\label{bpa4}
\end{equation}
and study \eqref{bpa3} perturbatively. As we will see, any motion of the perihelion can be seen only at the third order in perturbation. 

We note that if we substitute the leading order term (up to linear order in $Q_{p}$) from \eqref{bpa4} into \eqref{bpa3} we obtain
\begin{equation}
\left(\frac{du}{d\phi}\right)^{2} + u^{2} - \frac{(2k^{2} - 1) Q_{p}}{2\ell^{2}} u^{7-p} = \frac{(k^{2}-1)}{\ell^{2}}.\label{bpa5}
\end{equation}
From \eqref{bpa5} we can define the effective potential in which the particle moves as
\begin{equation}
V_{\rm eff} (u) = u^{2}  - \frac{(2k^{2} - 1) Q_{p}}{2\ell^{2}} u^{7-p},\label{bpa6}
\end{equation}
so that the extremum of the potential satisfies
\begin{equation}
V_{\rm eff}^{\prime} (u) = 2u \left(1 - \frac{ (7-p) (2k^{2} - 1) Q_{p}}{4\ell^{2}} u^{5-p}\right) = 0,\label{bpa7}
\end{equation}
where a prime denotes a derivative with respect to $u$. We note from \eqref{bpa7} that for $k^{2} > \frac{1}{2}$, Descartes rule of signs determines that the equation has only one positive root (physical solution) which is given by (we are considering $p\leq 6$)
\begin{equation}
u_{0} = \left(\frac{4\ell^{2}}{(7-p) (2k^{2} - 1) Q_{p}}\right)^{\frac{1}{5-p}}.\label{bpa8}
\end{equation}
On the other hand, for $k^{2} < \frac{1}{2}$, there is no positive root leading to the fact that bound motion is not possible in this case.  Therefore, we restrict only to the case $k^{2} > \frac{1}{2}$ where a physical root of \eqref{bpa7} is possible. To determine the stability of this solution, we note that
\begin{equation}
V_{\rm eff}^{\prime\prime} (u_{0}) = - \frac{(7-p) (5-p) (2k^{2}-1)Q_{p}}{2\ell^{2}} u_{0}^{5-p} = - 2 (5-p).\label{bpa9}
\end{equation}
This relation is very interesting because it shows that the stability of the motion depends on the type of branes that provide the gravitational background. For example, for $p < 5$, we see that $V_{\rm eff}^{\prime\prime} (u_{0}) < 0$ so that the motion is not stable and we cannot have any bound orbit. For  $p=5$, the second derivative vanishes so that motion is again unstable. Only for the case of $p = 6$, we have $V_{\rm eff}^{\prime\prime} (u_{0}) = 2 > 0$ so that stable motion is possible.  In fact, this seems distinctly different from the case of a Schwarzschild background where, as we have seen in subsection \ref{sgm}, no bound motion is possible for $d\geq 5$. However, in light of the discussions in \eqref{gm4}-\eqref{gm9}, we note from \eqref{bpa6} that with increasing dimension of the extended gravitational source, the fall off of the gravitational potential becomes slower and overtakes the centrifugal potential for $p=6$. This is the reason behind the stability of motion in the background of D$p$-branes.

Therefore, let us consider the particle motion for $k^{2} > \frac{1}{2}$ and $p=6$. In this case, \eqref{bpa5} as well as the equation following from this (we ignore the circular motion) can be written as
\begin{align}
& \left(\frac{du}{d\phi}\right)^{2} + u^{2} - \frac{(2k^{2} - 1) Q_{6}}{2\ell^{2}} u = \frac{(k^{2}-1)}{\ell^{2}},\notag\\
& \frac{d^{2}u}{d\phi^{2}} + u - \frac{(2k^{2} -1) Q_{6}}{4\ell^{2}} = 0.\label{bpa10}
\end{align}
This has the same structure as the motion of a point particle in a Schwarzschild background in $d=4$ and the solution satisfying the two relations in \eqref{bpa10} is given by
\begin{align}
u (\phi) & = \frac{(2k^{2}-1)Q_{6}}{4\ell^{2}} \left(1 + e_{1} \cos \phi\right),\notag\\
e_{1}^{2} & = 1 + \frac{16(k^{2}-1)\ell^{2}}{(2k^{2}-1)^{2}Q_{6}^{2}},\label{bpa11}
\end{align}
and we note that the bound motion is given by an ellipse with eccentricity $e_{1} < 1$ for $k^{2} < 1$ (in fact, we should have, in addition,  $\frac{16(1- k^{2})\ell^{2}}{(2k^{2}-1)^{2}Q_{6}^{2}} < 1$ for physical motion).  Therefore, we will restrict ourselves to $p=6, 1 > k^{2} > \frac{1}{2}$ (including the more stringent restriction given above) and in such a case, the perihelion and the aphelion are given by
\begin{equation}
r_{\sc\rm AH}^{(1)} = \frac{4\ell^{2}}{(2k^{2}-1) Q_{6}} \frac{1}{1-e_{1}},\quad r_{\sc\rm PH}^{(1)} = \frac{4\ell^{2}}{(2k^{2}-1) Q_{6}} \frac{1}{1 + e_{1}}.\label{bpa12}
\end{equation}

If we add the next order (second order in $Q_{6}$) contribution from \eqref{bpa4}, the two equations in \eqref{bpa10} generalize to
\begin{align}
& \left(\frac{du}{d\phi}\right)^{2} + \left(1 - \frac{Q_{6}^{2}}{8\ell^{2}}\right)u^{2} - \frac{(2k^{2} - 1) Q_{6}}{2\ell^{2}} u = \frac{(k^{2}-1)}{\ell^{2}},\notag\\
& \frac{d^{2}u}{d\phi^{2}} + \left(1 - \frac{Q_{6}^{2}}{8\ell^{2}}\right) u - \frac{(2k^{2} -1) Q_{6}}{4\ell^{2}} = 0,\label{bpa13}
\end{align}
which have the same structures as \eqref{bpa10} except for a small correction in one of the coefficients. The solution satisfying both these equations is given by
\begin{align}
u (\phi) & = \frac{(2k^{2}-1)Q_{6}}{4\ell^{2}}\left(1 - \frac{Q_{6}^{2}}{8\ell^{2}}\right)^{-1} \left(1 + e \cos \left(1 - \frac{Q_{6}^{2}}{8\ell^{2}}\right)^{\frac{1}{2}}\phi\right),\notag\\
e^{2} & = 1 + \frac{16(k^{2}-1)\ell^{2}}{(2k^{2}-1)^{2}Q_{6}^{2}}\left(1 - \frac{Q_{6}^{2}}{8\ell^{2}}\right),\label{bpa14}
\end{align}
and we see that the motion continues to be that of an ellipse with
\begin{equation}
r_{1} = r_{\sc\rm AH}^{(2)} = \frac{4\ell^{2}}{(2k^{2}-1) Q_{6}}\left(1 - \frac{Q_{6}^{2}}{8\ell^{2}}\right) \frac{1}{1-e},\quad r_{2} = r_{\sc\rm PH}^{(2)} = \frac{4\ell^{2}}{(2k^{2}-1) Q_{6}}\left(1 - \frac{Q_{6}^{2}}{8\ell^{2}}\right) \frac{1}{1 + e}.\label{bpa15}
\end{equation}

On the other hand, if we include the next order (cubic in $Q_{6}$) correction from \eqref{bpa4}, then \eqref{bpa3} takes the form
\begin{equation}
\left(\frac{du}{d\phi}\right)^{2} = \frac{(k^{2}-1)}{\ell^{2}} + \frac{(2k^{2}-1)Q_{6} u}{2\ell^{2}} - \left(1 - \frac{Q_{6}^{2}}{8\ell^{2}}\right) u^{2} - \epsilon u^{3},\quad \epsilon = \frac{Q_{6}^{3}}{16\ell^{2}}.\label{bpa16}
\end{equation}
This has exactly the same form (with some differences that we will elaborate on in the following) as that for a point particle in a Schwarzschild background in $4$ dimensions. We note that the locations of the perihelion and the aphelion are obtained by setting $\frac{du}{d\phi} = 0$ so that the right hand side leads to a cubic equation. If we denote the three roots as $u_{1}, u_{2}, u_{3}$, then it follows from the form of the cubic equation that
\begin{equation}
u_{1}+u_{2}+u_{3} = - \frac{1}{\epsilon}\left(1 - \frac{Q_{6}^{2}}{8\ell^{2}}\right),\label{bpa17}
\end{equation}
which is a large negative constant. Since we know that $u_{1},u_{2}$ are finite positive quantities (see \eqref{bpa15}), it follows that $u_{3}$ must be large and negative and the motion is bounded by $u_{1}\leq u \leq u_{2}$. Following the standard method, we can now determine from \eqref{bpa16}
\begin{equation}
\frac{d\phi}{du} \simeq \frac{1 - \frac{\epsilon}{2} (u+ u_{1} + u_{2})}{\sqrt{(u-u_{1})(u_{2}-u)}},\label{bpa18}
\end{equation}
which can be integrated to give
\begin{equation}
|\delta\phi| = \int\limits_{u_{1}}^{u_{2}} du\,\frac{1 - \frac{\epsilon}{2} (u+u_{1}+u_{2})}{\sqrt{(u-u_{1})(u_{2}-u)}} = \pi \left(1 - \frac{\epsilon}{4} (u_{1} + u_{2})\right).\label{bpa19}
\end{equation}
It follows, therefore, that
\begin{equation}
\Delta\phi = 2 |\delta\phi| - 2\pi = - \frac{\pi\epsilon}{2} (u_{1}+u_{2}) = - \frac{\pi Q_{6}^{3}}{32\ell^{2}} \left(\frac{1}{r_{1}} + \frac{1}{r_{2}}\right).\label{bpa20}
\end{equation}
Namely, at the end of a period, the perihelion does not come back to its old location, rather, there is regression in its position. This is interesting because in the standard Schwarzschild case, the perihelion advances at the end of a period. This difference can be understood from the fact that while in the Schwarzschild case, the relativistic gravitational interaction is attractive, here the cubic term in \eqref{bpa16} leads to a repulsive interaction which slows the particle and hence leads to a regression. Of course, the motion of the perihelion is extremely tiny as in the case of the Schwarzschild background in $4$ dimensions. However, this qualitative difference in the behavior of a brane background from that of the Schwarzschild background is most certainly interesting and any possible test for the presence or absence of this would be quite remarkable. 

\subsection{Motion of light rays}

Let us next discuss the motion of light rays (photons) in the background of a stack of D$p$ branes. As we have discussed earlier, in this case $d\tau^{2} = 0$ and the trajectory has to be parametrized by an affine parameter, say $\lambda$. As a result, the line element leads to
\begin{equation}
H^{-\frac{1}{2}} (r) \frac{\dot{t}^{2}}{\dot{\phi}^{2}} - H^{\frac{1}{2}} (r) \left(\frac{\dot{r}^{2}}{\dot{\phi}^{2}} + r^{2}\right) = 0,\label{blr1}
\end{equation}
where dots denote derivatives with respect to the affine parameter $\lambda$. Using the definitions in \eqref{b2} we can write this relation as
\begin{equation}
\left(\frac{dr}{d\phi}\right)^{2} + r^{2} - \frac{k^{2}}{\ell^{2}} r^{4} H (r) = 0.\label{blr2}
\end{equation}
Using the new variable $r = \frac{1}{u}$, this equation can be written as
\begin{equation}
\left(\frac{du}{d\phi}\right)^{2} + u^{2} - \frac{k^{2} Q_{p}}{\ell^{2}} u^{7-p} = \frac{k^{2}}{\ell^{2}},\label{blr3}
\end{equation}
which can be compared with \eqref{lr4}. If we differentiate this relation with respect to $\phi$, we obtain
\begin{equation}
\frac{du}{d\phi} \left(\frac{d^{2}u}{d\phi^{2}} + u - \frac{k^{2} (7-p) Q_{p}}{2\ell^{2}} u^{6-p}\right) = 0,\label{blr4}
\end{equation}
which leads to
\begin{equation}
\frac{du}{d\phi} = 0,\quad \text{or,}\quad \frac{d^{2}u}{d\phi^{2}} + u - \frac{k^{2} (7-p) Q_{p}}{2\ell^{2}} u^{6-p} = 0.\label{blr5}
\end{equation}
The first equation in \eqref{blr5} describes any possible circular orbit for photons which we have already discussed in \eqref{bco11}. Therefore, let us look at the more general second equation in \eqref{blr5}. 

We note that the second equation in \eqref{blr5} can be written as
\begin{equation}
\frac{d^{2}u}{d\phi^{2}} + u = \epsilon u^{6-p},\label{blr6}
\end{equation}
where we have identified $\epsilon = \frac{k^{2} (7-p) Q_{p}}{2\ell^{2}}$, which is a very small parameter. Comparing with \eqref{lr6a} we see that  \eqref{blr6} has the same structure with the identification $d = 8-p$ and the analysis in \eqref{lr7}-\eqref{lr19} can be carried through completely in a similar manner. The deflection in the path of a light ray due to the presence of a stack of D$p$ branes can now be determined to be
\begin{equation}
\Delta \simeq \left\{\begin{matrix}
\frac{k^{2} Q_{p} u_{\rm min}^{5-p}}{\ell^{2} (5-p)} \prod\limits_{n=1}^{3- \frac{p}{2}} \left(1 + \frac{3}{5-p-2n}\right), & p\ \text{even},\\
\frac{3\pi k^{2} Q_{p} u_{\rm min}^{5-p}}{2\ell^{2} (5-p)} \prod\limits_{n=1}^{2- [\frac{p}{2}]} \left(1 + \frac{3}{5-p-2n}\right), & p\ \text{odd}
\end{matrix}\right. .\label{blr7}
\end{equation}

\section{Conclusion}

In this paper, we have systematically analyzed the motion of a point particle (massive and massless) in the gravitational background of a stack of D$p$-branes in $10$ dimensions. Such extended objects exist in higher dimensional theories of gravity such as the string theory and, therefore, it is quite natural to study the effect of such gravitational sources on test particles and to compare and contrast them with the effects produced by a Schwarzschild background. For this purpose, we have also systematically analyzed the effects of the Schwarzschild background on test particles in higher dimensions (some of which may be available in a scattered form) and then carried out the similar analysis for the D$p$-brane backgrounds. In both cases, we have studied the radial fall of a massive particle, the motion of a particle (massive and massless) in a circular orbit as well as the general motion of a massive particle. In addition, we have also studied the question of the deflection in the path of photons in such  backgrounds and we summarize our results in the following.

The analysis of the radial fall of a massive particle in the Schwarzschild background leads to the well known fact that the (coordinate) speed of the particle falling initially from rest at $r_{0}$ keeps on increasing up to a point and then starts to decrease until it vanishes at the Schwarzschild radius (in $d$ dimensions). In the background of the D$p$-branes, the particle falls from rest and gains speed up to a certain point and then the speed decreases, but vanishes only at the origin. It is interesting that independent of the number of dimensions of space-time and the nature of the gravitational source (namely, Schwarzschild or D$p$-branes), the maximum speed that the particle can have (for large $r_{0}$) is given by $v_{\rm max} = \frac{2}{3\sqrt{3}}$.

In a higher dimensional Schwarzschild background, circular orbits for a massive point particle are possible if the radius of the orbit satisfies 
\begin{equation}
r > \left(\frac{(d-1)}{2}\right)^{\frac{1}{d-3}}\, r_{\sc S}^{(d)},
\end{equation}
where $d$ denotes the number of space-time dimensions. For a background of D$p$-branes, on the other hand, circular orbits are possible without any restriction on $r$ for $p\geq 3$ while for $p < 3$, a circular orbit requires the radius to satisfy
\begin{equation}
r > \left(\frac{(3-p)Q_{p}}{4}\right)^{\frac{1}{7-p}}.
\end{equation}
In all the cases, we have shown that for orbits with large $r$, a generalization of Kepler's law holds, namely, the square of the period is proportional to $r^{\alpha}$ where $\alpha$ denotes the number of transverse spatial dimensions associated with the given gravitational source.

For a massive particle in a Schwarzschild background there is no noncircular bound orbits for $d\geq 5$ (so that the motion of the perihelion of an elliptic orbit is not a meaningful question). This was the main result in the work of \cite{tangherlini} and this can be understood qualitatively from the fact that for $r\gg r_{\sc S}$, the attractive gravitational potential falls off much faster than the repulsive centrifugal potential leading to instability in motion. In contrast, the background of D$p$-branes does allow bound elliptic motion for $p=6$. The reason for this lies in the fact that as $p$ increases, the fall off in the gravitational potential becomes slower and overtakes the repulsive centrifugal potential precisely for $p=6$. Consequently, one can study questions such as the perihelion motion in this case. We find that when the relativistic corrections are included, the perihelion of the ellipse regresses (as opposed to the conventional advancement) and this can be traced to the fact that the gravitational potential resulting from the D$p$-brane metric leads to a repulsive potential at this order contrasted with the attractive potential resulting from a Schwarzschild background. This is a significant difference between the two backgrounds and if this can be tested that would indeed be quite remarkable. We note that like the conventional perihelion motion, this effect is extremely tiny.

For massless particles (photons), the Schwarzschild background permits a circular orbit for
\begin{equation}
r = \left(\frac{(d-1)}{2}\right)^{\frac{1}{d-3}} r_{\sc S}^{(d)},
\end{equation}
which lies very close to the Schwarzschild radius in any space-time dimension. For the background of D$p$-branes, on the other hand, a circular orbit for photons is allowed only for $p < 3$ at the radius
\begin{equation}
r = \left(\frac{(3-p)Q_{p}}{4}\right)^{\frac{1}{7-p}},
\end{equation}
whereas there are no circular orbits for photons for branes with $p\geq 3$. Bending of light in the presence of these two backgrounds can also be calculated and are similar in nature.
\bigskip

\noindent{\bf Acknowledgments}
\medskip

This work was supported in part  by US DOE Grant number DE-FG 02-91ER40685 (A.D. and L.G.) as well as by grant No. 2009/37/24 BRNS, India (A.B. and S.P.).

\end{document}